\documentclass{article}

\usepackage{microtype}
\usepackage{graphicx}
\usepackage{subfigure}
\usepackage{amsmath}

\usepackage{subcaption}
\usepackage{booktabs} 

\usepackage{hyperref}
\usepackage{enumitem}

\usepackage{caption}

\usepackage[accepted]{mlsys2024}

\mlsystitlerunning{ESPN: Memory-Efficient Multi-Vector Information Retrieval}

\begin{document}


\twocolumn[
\mlsystitle{ESPN: Memory-Efficient Multi-Vector Information Retrieval}



\mlsyssetsymbol{equal}{*}

\begin{mlsysauthorlist}
\mlsysauthor{Susav Shrestha}{TAMU}
\mlsysauthor{Narasimha Reddy}{TAMU}
\mlsysauthor{Zongwang Li}{Samsung}
\end{mlsysauthorlist}

\mlsysaffiliation{TAMU}{Department of Electrical and Computer Engineering, Texas A\&M University, Texas, USA}
\mlsysaffiliation{Samsung}{Samsung Semiconductor, San Jose, California, USA}

\mlsyscorrespondingauthor{Susav Shrestha}{sls7161@tamu.edu}

\mlsyskeywords{Machine Learning, MLSys}

\vskip 0.3in

\begin{abstract}
Recent advances in large language models have demonstrated remarkable effectiveness in information retrieval (IR) tasks. While many neural IR systems encode queries and documents into single-vector representations, multi-vector models elevate the retrieval quality by producing multi-vector representations and facilitating similarity searches at the granularity of individual tokens. However, these models significantly amplify memory and storage requirements for retrieval indices by an order of magnitude. This escalation in index size renders the scalability of multi-vector IR models progressively challenging due to their substantial memory demands. We introduce Embedding from Storage Pipelined Network (ESPN) where we offload the entire re-ranking embedding tables to SSDs and reduce the memory requirements by \(5-16\times\). We design a software prefetcher with hit rates exceeding 90\%, improving SSD based retrieval up to \(6.4\times\), and demonstrate that we can maintain near memory levels of query latency even for large query batch sizes. The code is maintained at \href{https://github.com/susavlsh10/ESPN-v1/}{https://github.com/susavlsh10/ESPN-v1/}

\end{abstract}
]



\printAffiliationsAndNotice{ }  

\section{Introduction}
\label{submission}

Recent advances in natural language processing and the emergence of large language models (LLMs) have led to a substantial uplift in Information Retrieval (IR) systems \cite{NIPS2017_3f5ee243, DBLP:journals/corr/abs-1910-14424, DBLP:journals/corr/abs-2004-04906}. These improvements in retrieval quality come at the cost of an ever-growing appetite for computational and memory resources. While advancements in GPU and TPU architectures have largely addressed the computational requirements for efficient inference, the challenge of expanding memory remains prohibitively expensive, necessitating the scaling up of hardware resources.

Modern neural IR systems leverage these fine-tuned LLMs to encode text documents into dense vectors or embeddings, effectively capturing their textual essence \cite{DBLP:journals/corr/abs-2005-00181, devlin-etal-2019-bert, DBLP:journals/corr/abs-1802-05365}. In the landscape of embedding based retrieval, single vector systems streamline the retrieval process by encapsulating each text document within a single dense vector, facilitating similarity computations through methods like cosine similarity or dot product. Advancing this paradigm, late interaction models like ColBERT encode documents at the granularity of tokens, resulting in multi-vector representations \cite{10.1145/3397271.3401075}. The relevance between the query and the documents are modeled by summing the maximum similarity between each query vector and the entirety of document vectors.

The late interaction models bypass the information bottleneck of having to encode complex query document relationship within a single dot product by decomposing relevance modeling into token-level computations. However, this heightened level of expressiveness comes with its own set of trade-offs. Late interaction systems introduce a memory and storage footprint that surpasses that of single-vector and lexical models by an order of magnitude. For example, the index size of the ColBERTv1 was \(210\times\) larger than the index size of traditional lexical retrievers like BM25. There has been substantial work in this area to decrease the index size through various compression and reduction techniques while improving retrieval scores \cite{santhanam-etal-2022-colbertv2, 10.1145/3511808.3557325, 10.1145/3511808.3557367, cohen-etal-2022-sdr}. However, the index size of these multi-vector models is still \(26-34\times\) larger than the lexical retrievers as shown in Table \ref{tab:motivationtable}. 

\begin{figure}
    \centering
    \includegraphics[width=0.9\linewidth]{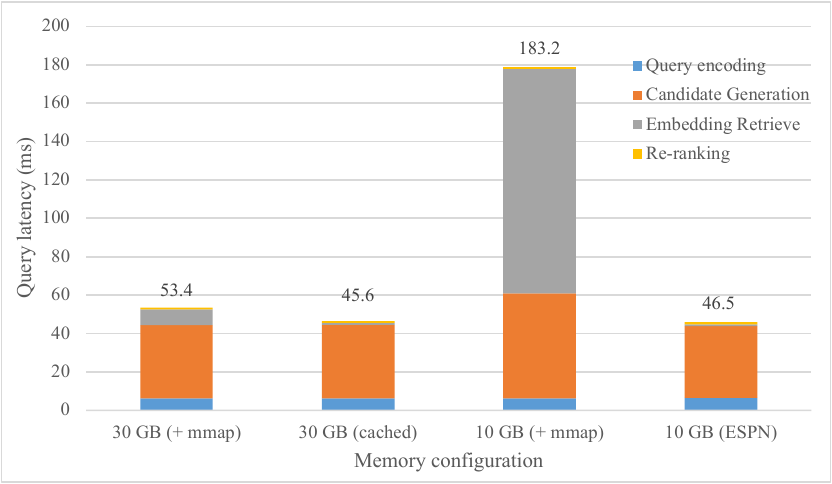}
    \caption{End-to-end query latency breakdown of ColBERTer in MS-MARCO v1 dataset with different memory configurations with index size of 18.8 GB. }
    \label{fig:motivation}
\end{figure}

The index size of the multi-vector models is of the order \(O(Ntdb)\) and is determined by 4 variables: 1) the number of documents, \textit{N}; 2) the number of token/word vectors saved per document, \textit{t}; 3) the dimensionality of the vector, \textit{d}; 4) the number of bytes per dimension, \textit{b}. Furthermore, to facilitate efficient retrieval, these systems also require an (approximate) nearest neighbor index or a traditional lexical index as a candidate generator or a first stage retriever to quickly select top-K documents for re-ranking. Therefore, the total index size of these system has the order of \(O(NI+Ntdb)\) where \(I\) is the average size of index per document in the candidate generator. 

Many recent endeavors in neural information retrieval optimize query latencies by storing document embedding vectors in system memory, allowing for fast lookups during the document ranking phase (Table \ref{tab:motivationtable}). However, this design's scalability is challenged when dealing with larger databases that surpass the limitations of the available system memory. As the size of the database grows, multi-vector retrievers run into scalability and memory limits much faster than single-vector and lexical retrievers because of their massive index size. Serving large-scale databases near memory capacity while employing conventional methods like memory-mapping (mmap) significantly degrades the query latency of late interaction models, as illustrated in Figure \ref{fig:motivation}. This surge in index size poses a distinct challenge to the scalability of these systems for managing large web-scale databases. 

While one solution involves scaling up server infrastructure to meet increased memory demands, a more cost-efficient alternative capitalizes on the growing capabilities of solid-state drives. These drives offer terabytes of storage per unit and boast low latencies, making them an effective choice for embedding retrieval. In this work, we study how we can design efficient multi-vector retrieval systems which can efficiently scale to larger datasets. We show that we can completely offload the bag-of-word (BOW) re-ranking embedding index \(O(Ntdb) \) to SSDs and apply our software prefetcher to maintain near memory level of query latency. We present the Embedding from Storage Pipelined Network (ESPN), a system that combines GPU-based prefetching and early re-ranking with concurrent CPU-driven nearest neighbor search. We implement the word embedding retrieval using Nvidia’s GPUDirect Storage which allows the vectors to bypass the host and directly transfer data to the GPU memory. Furthermore, we design a simple nearest neighbor prefetcher which hides the latency from storage in the critical path by overlapping I/O with meaningful computation. Our results demonstrate that our adaptable prefetching mechanism consistently achieves an effective hit rate that can exceeds 90\%. This flexibility enables our system to sustain low query latency, even when handling multiple concurrent queries on an SSD. Finally, we study bandwidth efficient solutions that allows SSD based solutions with ESPN to scale to large query batch sizes with marginal 0.3-0.7\% quality degradation. This paper makes the following contributions.

1. We identify scalability issues and the limitation of traditional O/S techniques like swapping for neural IR systems. ESPN is \(3.9\times\) faster in end-to-end query latency compared to mmap when operating near memory capacity.

2. We design a scalable embedding retrieval architecture for multi-vector IR by offloading re-ranking embedding tables to SSDs, reducing the memory requirements for fast retrieval by 5-16x depending on the quantization factor.

3. We identify a new opportunity to overlap storage latency with computation by designing a simple and flexible approximate nearest neighbor based prefetcher and improve SSD access latency up to \(6.4\times\)

4.	We explored bandwidth efficient storage solutions to multi-vector IR and demonstrated that partial re-ranking with marginal 0.3-0.7\% quality degradation can reduce the bandwidth requirements by 8-16\(\times\) and allows SSD based retrieval to scale to large batch queries.

\section{Background}

This section describes Neural Information Retrieval Systems (\ref{sec:NIRS}), the index size and memory requirements for efficient search (\ref{sec:Index}), and the memory hierarchy for storing retrieval indices (\ref{sec:MemoryHierarchy})

\subsection{Neural Information Retrieval Systems} \label{sec:NIRS} 
\begin{figure*}[h]
    \centering
    \includegraphics[width=0.9\linewidth]{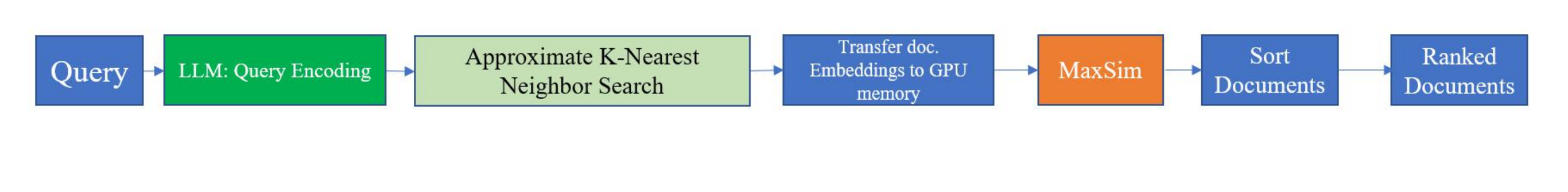}
    \caption{Conventional Neural Multi-Vector Retrieval Pipeline}
    \label{fig:ConventionalMethod}
\end{figure*}

Conventional lexical retrievers have long formed the backbone of information retrieval, relying on keyword-based matching to retrieve relevant documents \cite{10.1561/1500000019}. However, their efficacy can be constrained by the limitations of keyword matching, often leading to challenges in handling synonyms, context, and nuanced query formulations. Recent works have demonstrated that fine-tuning pre-trained large language models like BERT can effectively address limitations in semantic mismatch and achieve state-of-the-art results in several retrieval tasks \cite{devlin-etal-2019-bert, 10.1145/3397271.3401075}. 

Modern neural IR systems have shifted towards using these models for offline document indexing, recognizing the computational complexity and time constraints inherent in ranking thousands of documents during query processing. The offline indexing process generates document embedding vectors which encapsulate the contextual and semantic meaning of the text documents. Single vector models generate a single dense representation of the text documents and compute the similarity with the query using a dot product operation \cite{Formal2021SPLADEVS, ren-etal-2021-rocketqav2}. Moreover, multi-vector models adopt a late interaction architecture where the documents are indexed as matrices of word/token embedding vectors and the similarity with the query matrix is computed by a MaxSim operation as shown in equation (\ref{eq:eq1}) \cite{santhanam-etal-2022-colbertv2, 10.1145/3511808.3557367, li2022citadel}. In equation (\ref{eq:eq1}), \(S_{q,d}\) represents the similarity score of the current document \textit{d}, with query \textit{q} where \(E_q\) and \(E_d\) represents the query and the document embedding matrices generated using the language model.

\begin{equation}
S_{q,d} = \sum_{i\in [|E_q|]} \max_{j\in [|E_d|]} \left( E_{q_i} \cdot E_{d_j}^T \right) \label{eq:eq1}
\end{equation}

The MaxSim operation calculates the maximum similarity score between each query token and all document tokens, resulting in a fine-grained token-level representation of query-document similarity. The MaxSim operator, which involves computationally intensive multiple matrix multiplication operations between various documents embeddings and the query, is typically processed on GPUs within these systems. Computing similarity scores across the entire database of documents proves to be prohibitively expensive, leading these systems to embrace a retrieve and re-rank architecture. The retrieval stage is tasked with efficiently producing a list of candidate documents, commonly achieved through methods such as nearest neighbor searches or lexical retrievals. The re-ranking stage utilizes a scoring function to sort the documents based on their scores, resulting in the generation of a ranked document list. The retrieval process is demonstrated in figure \ref{fig:ConventionalMethod}. 
In the discussions below, we will measure the effectiveness of the retrieval system using the MRR@K and Recall@K metrics, the standard metric used in many IR datasets. MRR@K (Mean Reciprocal Rank at K) is an information retrieval metric that measures the average reciprocal of the rank at which the first relevant document is found among the top K ranked documents for a set of queries, prioritizing the position of the relevant document. Recall@K evaluates how effectively a retrieval system can find relevant documents within the top K positions of the ranked list for a set of queries, expressed as a percentage of relevant documents found.

\subsection{Index size and memory requirements for efficient retrieval}\label{sec:Index} 

Table \ref{tab:motivationtable} shows the index size and the retrieval latencies on the MS-MARCO v1 dataset with 8.8M passages as reported in various papers in this field. The table highlights the differences in retrieval quality, index size, and latency of various models. Compared to single vector, lexical and impact-based retrievers, multi-vector models achieve higher MRR@10 scores at the cost of an expanded index size. It is important to note that all these retrieval systems compute the query latency with the entire index cached in memory.

\begin{table}[htbp]
\centering
\caption{In-domain index size, latency and performance on the development set of MS MARCO passage ranking dataset. We copy the results from their respective papers: \cite{mackenzie2021wacky, Cheriton2019FromDT,Formal2021SPLADEVS, lin2021brief, 10.1145/3404835.3463030, 10.1145/3397271.3401075, gao-etal-2021-coil, 10.1145/3511808.3557367, santhanam-etal-2022-colbertv2, 10.1145/3511808.3557325}. We report the latencies from \cite{mackenzie2021wacky} for non multi-vector models.}
\resizebox{\columnwidth}{!}{%
\begin{tabular}{lcccccc}
\toprule
\textbf{Model} & \textbf{Index size} & \textbf{Factor} & \textbf{Query Latency} & \textbf{MRR@10} & \textbf{R@1K} \\
\midrule
\multicolumn{6}{l}{\textbf{Lexical, single vector, and impact models}} \\
BM25 (PISA)  & 0.73 GB & $\times1$ & 8 ms & 0.194 & 0.868 \\
docT5query & 1.04 GB & $\times1.4$ & 11.9 ms & 0.277 & 0.947 \\
Spladev2 & 4.3 GB & $\times5.9$ & 220 ms & 0.369 & 0.979 \\
UniCOIL-Tok  & 1.4 GB & $\times1.9$ & 37 ms & 0.352 & - \\
DeepImpact  & 1.56 GB & $\times2.1$ & 19.4 ms & 0.326 & 0.948 \\
\addlinespace 
\multicolumn{6}{l}{\textbf{Multi vector models}} \\
ColBERTv1  & 154 GB & $\times210.9$ & 54.3 ms & 0.367 & 0.968 \\
COIL (Dim 768, 32)  & 54.7 GB & $\times74.9$ & 41 ms & 0.355 & 0.963 \\
ColBERTer (Dim 32)  & 18.8 GB & $\times25.8$ & 51 ms & 0.387 & 0.961 \\
ColBERTv2 & 24.6 GB & $\times33.7$ & 259.6 ms & 0.397 & 0.983 \\
ColBERTv2 (PLAID) & 21.6 GB & $\times29.6$ & 38.4 ms & 0.398 & 0.975 \\
\bottomrule
\end{tabular}%
}
\label{tab:motivationtable}
\end{table}

The success of ColBERTv1 opened the doors for multi-vector retrieval, but it also increased the index size by 210x when compared to BM25. Subsequent works have attempted to decrease the index size and memory requirements through dimension reduction, quantization, filtering and compression strategies \cite{DBLP:journals/corr/abs-2012-15156, 10.1145/3459637.3482358, cohen-etal-2022-sdr, liu2022dimension}. ColBERTv2 and ColBERTer incorporate many of these strategies and reduce the index size by 6.2x and 8.2x respectively when compared to ColBERTv1. Even with these improvements, the overall index size of these multi-vector models is still 26-34x larger than the BM25 baseline index. The index size becomes an even bigger issue when the size of the dataset increases. Table \ref{tab:motivationv2} shows the index size for MS-MARCO v2 dataset with 138 million passages. 

\begin{table}[htbp]
\centering
\caption{Out of domain retrieval scores, index size and query latency in MS MARCO v2 dataset. We copy the results for ColBERTv2 and PLAID from \cite{10.1145/3511808.3557325}. * suggests our test results with publicly available model, higher recall scores could potentially be achieved at higher latencies.}
\resizebox{\columnwidth}{!}{%
\begin{tabular}{lccccccc}
\toprule
\textbf{Model} & \textbf{Index size} & \textbf{MRR@100} & \textbf{R@1K} & \textbf{Latency (ms)} \\
\midrule
BM25 & 11 GB & 0.087 & 69.3 & - \\
ColBERTer & 290 GB* & 0.15* & 71.8* & CPU + GPU 85.7 \\
ColBERTv2 & 246 GB & 0.18 & 88.1 & GPU (OOM) CPU 5228.5 \\
ColBERTv2 (PLAID) & 202.2 & 0.18 & 85.7 & GPU (OOM) CPU 251.3 \\
\bottomrule
\label{tab:motivationv2}
\end{tabular}%
}

\end{table}
When considering this larger dataset, it becomes apparent that despite employing various strategies to minimize the index size, the multi-vector models still entail a significant memory and storage footprint. Although larger disk space usage may not be a primary cost concern, maintaining numerous pre-computed representations in memory, as often required for low latency systems, does result in a substantial increase in hardware expenses. This increase in index size raises immediate concerns regarding scalability, particularly when transitioning to web-scale applications. 

In the next section, we focus on one of the multi-vector models and discuss SSD based solutions for IR. ColBERTv2 adopts a sophisticated decompression pipeline which tightly couples residual decompression with ANN centroids. To facilitate straightforward comparisons, we utilize the publicly available ColBERTer model which follows the retrieve and re-rank approach.

\subsection{Memory hierarchy for storing retrieval indices} \label{sec:MemoryHierarchy} 
\label{author info}

\begin{figure}
    \centering
    \includegraphics[width=0.8\linewidth]{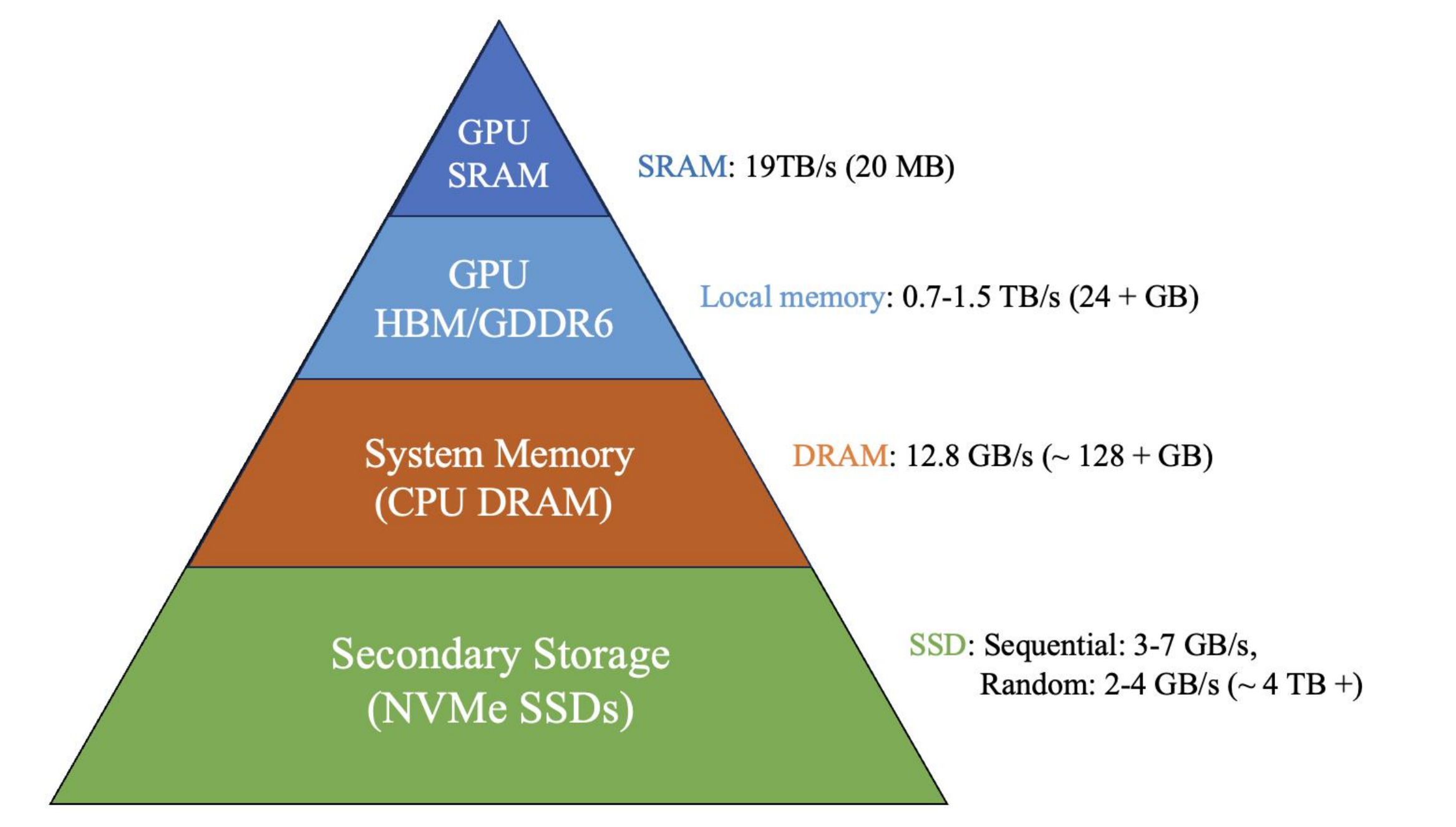}
    \caption{Memory Hierarchy for storing Retrieval Indices.}
    \label{fig:Indexhierarchy}
\end{figure}

Index and embedding tables can be stored or cached at various levels of the memory hierarchy offering different sizes and speeds, with smaller memory being faster and more expensive. The hierarchy for storing retrieval indices is shown in figure \ref{fig:Indexhierarchy}. Storing retrieval indices in GPU memory would be ideal as it provides sufficient memory bandwidth for search, but GPU HBM is often a very limited resource. Scaling GPU memory requires linear scaling of GPUs which can become very expensive very fast. With the increasing adoption of retrieval augmented generation (RAG) techniques \cite{10.5555/3495724.3496517} and LLMs for search, GPU memory is often used for caching billions of LLM parameters . 

Many recent works, as shown in Table \ref{tab:motivationtable}, in neural IR achieve low query latency by caching the entire retrieval index in system memory. Storing a part of these indices in Solid State Drives (SSDs), which provide lower latencies and higher throughput compared to traditional drives, is often a more cost-effective and scalable solution. However, a naive SSD based solution can introduce storage-based latency in the critical path of the query which might not be desirable for low latency systems. SSD based solutions may also encounter challenges when handling queries in large batches, demanding higher bandwidth for efficient search. Furthermore, it is crucial to exercise caution when designing storage-based solutions as conventional software stacks like mmap can introduce significant software overhead for random reads.

Conventional systems address memory constraints using techniques like memory mapping (mmap) and swap space. Mmap integrates files into a process's virtual memory, allowing on-demand retrieval and caching in memory. Swap space extends physical memory and is used when physical memory is exhausted, avoiding crashes due to low memory. While these methods work with minimal code changes, they introduce software overhead and can be inefficient when accessing data not in memory. A single query thread can access thousands of different memory locations, and employing blocking page fault handling to access each page from storage adds a significant amount of software overhead, leading to inefficient utilization of SSD bandwidth. The inefficiencies of mmap have previously been discussed in the context of database systems \cite{crotty22-mmap}.
We discuss the results and the overheads of using mmap and swap space in section 5.

\section{SSD based Multi-Vector Information Retrieval }

Solid-state drives (SSDs) excel in random read operations, outperforming traditional hard drives by an order of magnitude. SATA SSDs typically deliver around 100K random IOPS, while NVMe SSDs featuring the newer PCIe 4.0 interface can deliver up to 1M random IOPS \cite{Samsung990Pro}. To attain these numbers, it is important to utilize asynchronous I/O to efficiently populate the internal SSD queue and request multiple blocks simultaneously. Additionally, in the context of accelerator and GPU-based systems, traditional file I/O necessitates redundant data transfers from SSD to CPU memory and then from CPU memory to GPU memory. Nvidia's new I/O infrastructure, such as GPUDirect Storage (GDS) \cite{NVIDIADocs}, facilitates direct peer-to-peer data transfers from SSD to GPU memory. However, there is still a large gap in performance between DRAM and SSD both in terms of latency (1000x) and random bandwidth (3-6x), and a naïve solution would certainly degrade the query processing latency significantly.  

\subsection{Index structure in Retrieve and Re-rank IR models }

ColBERTer jointly fine-tunes a distilBERT \cite{Sanh2019DistilBERTAD} model such that it generates a single-vector and a multi-vector representation that can both be used for retrieval. The single vector representation (CLS vector) is much smaller (7.7x uncompressed) than the multi-vector representation (BOW vector) and is used to train an approximate nearest neighbor index which is used for candidate generation. The nearest neighbor list from the candidate generation is used to access the bag-of-word (BOW) embeddings for the re-ranking stage. The re-ranking applies the MaxSim operator (equation \ref{eq:eq1}) and combines the BOW scores (MaxSim) with the CLS scores (dot product) from the candidate generation step using a learned scaling factor. The documents are ranked using this aggregate score to generate the final list of ranked documents. Table 3 provides a summary of ColBERTer's index size, categorized into CLS and BOW vectors.

\begin{table}[htbp]
    \centering
    \caption{Index breakdown in ColBERTer}
    \resizebox{\columnwidth}{!}{%
    \begin{tabular}{lcccccc}
        \toprule
        \textbf{Dataset} & \textbf{\# Passages} & \textbf{\# Tokens} & \textbf{\# Queries} & \textbf{CLS (GB)} & \textbf{BOW (GB)} \\
        \midrule
        MS MARCO v1 & 8.8M & 597.9M & 6980 & 2.1 & 16.8 \\
        MS MARCO v2 & 138.4M & 9.4B & 3903 & 34.6 & 255.4 \\
        \bottomrule
    \end{tabular}
    }
    \label{tab:ColBERTerIndex}
\end{table}

The numbers presented in table 3 represent uncompressed and unquantized 16-bit vectors from ColBERTer. The CLS embeddings use 128 dimensional vectors per passage and BOW embeddings use 32 dimensional vectors per token as in the original paper. In this work, we propose to partition the candidate generation index in memory and offload the BOW embeddings retrieval to storage using our novel strategies. We focus our research on reducing the access latency from SSDs and reduce the memory required for efficient search.

\begin{figure*}[h]
    \centering
    \includegraphics[width=1\linewidth]{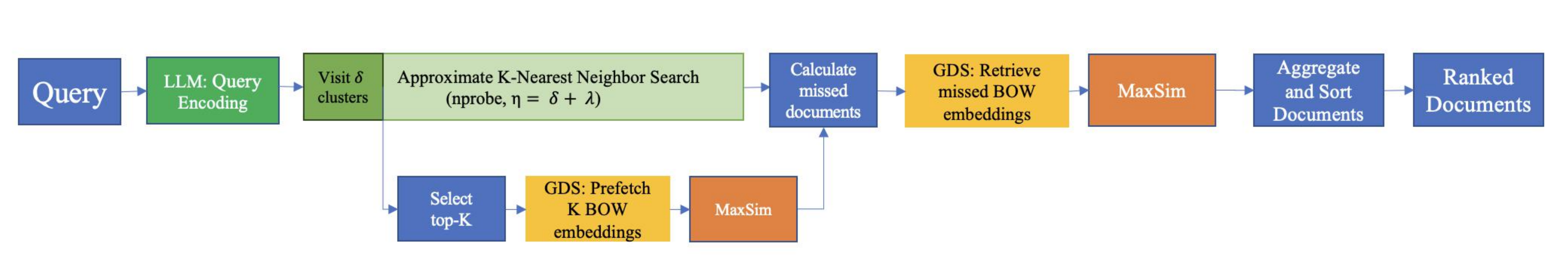}
    \caption{ESPN Retrieval Architecture}
    \label{fig:ESPN}
\end{figure*}

\section{Embedding from Storage Pipelined Network Architecture}
We present Embedding from Storage Pipelined Network (ESPN), a system that combines GDS based retrieval with software prefetching and minimizes the latency from SSD in the critical path of the query. Figure \ref{fig:ESPN} illustrates ESPN retrieval pipeline which consists of GPUDirect Storage based embedding retrievals, software prefetching, and early reranking. We partition the candidate generation (ANN) index and embedding table metadata (offsets) in CPU memory, re-ranking BOW embedding tables on the SSD, and utilize the GPU memory for caching the LLM parameters. This design is particularly useful for RAG systems that requires storing LLMs with billions of parameters in GPU memory. 

\subsection{Embedding retrieval optimizations using Nvidia GPUDirect Storage}
\label{final author}
We build our embedding retrieval system on top of Nvidia’s GPUDirect Storage batch APIs which enables asynchronous and direct data transfers from SSD to GPU memory. A similar retrieval architecture can also be built using traditional asynchronous I/O for systems that do not support GPUDirect Storage. In our effort to minimize the number of blocks being read per query, we strategically align the CLS embeddings and BOW embeddings together within the embedding binary file. This optimization proves especially beneficial when the collective size of the embeddings falls below I/O block size, as it reduces the required number of blocks from 2 to just 1 per document. This scenario is a common occurrence in both ColBERTer and ColBERTv2 because of their adoption of compression and reduction strategies. While the retrieval of CLS embeddings may not be required for ANN based candidate generation, this architectural choice paves the way for the streamlined integration of the ColBERTer model with other lexical and impact-based retrievers. 

\subsection{Approximate nearest neighbor prefetching}

\begin{figure}
    \centering 
    \includegraphics[width=0.9\linewidth]{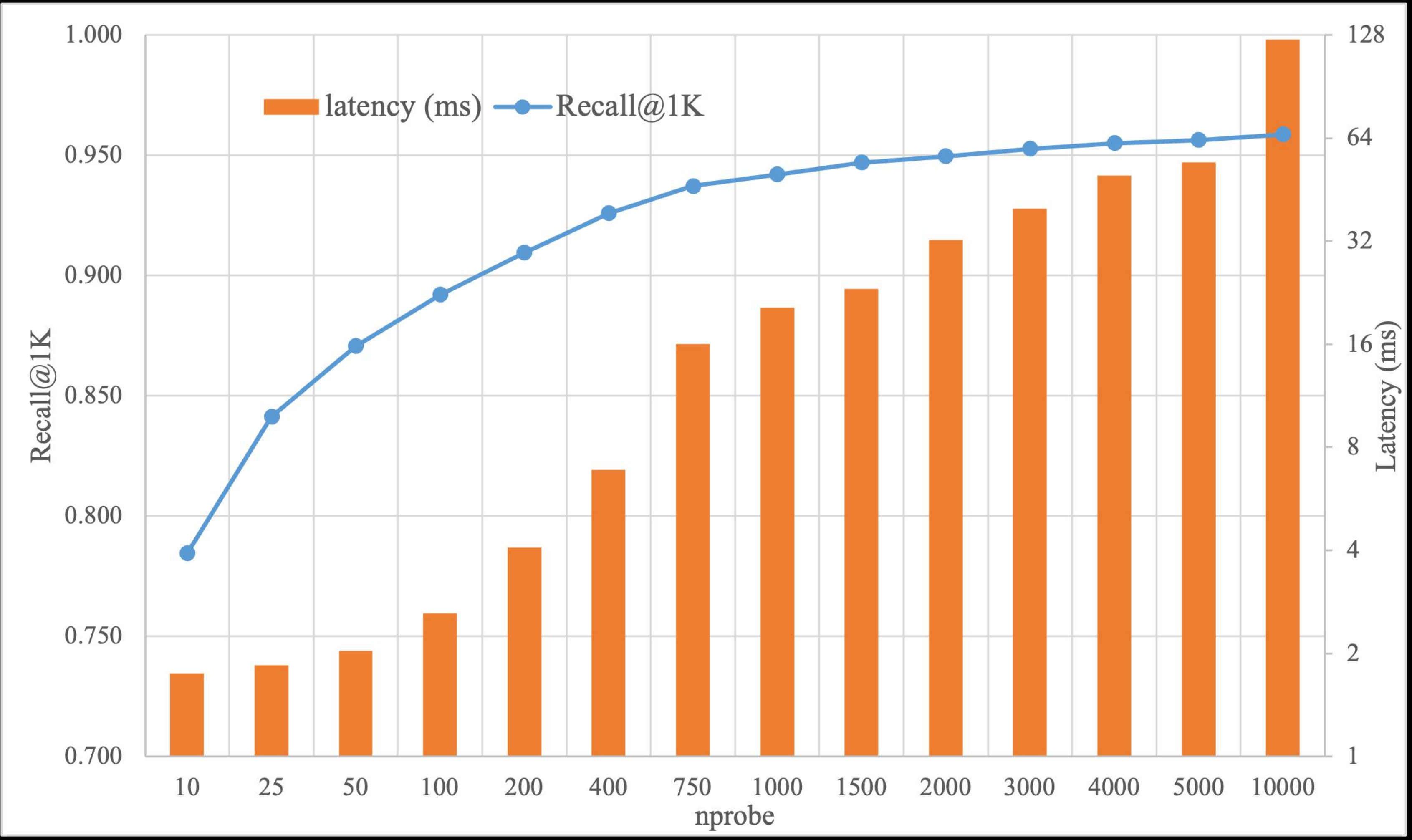} 
    \caption{Recall@1K vs nprobe with ColBERTer in MS MARCO v1 dataset using Faiss. nlist = \(2^{15}\)} 
    \label{fig:Recallvsnprobe} 
\end{figure}

Approximate nearest neighbor (ANN) algorithms such as Faiss and DiskANN \cite{DBLP:journals/corr/JohnsonDJ17, subramanya2019diskann} form the backbone of many vector databases. In this work, we focus on memory efficiency and thus build our system around inverted file (IVF) based search as HNSW based algorithms inflates the index memory size \cite{8594636}.
IVF based algorithms employ advanced techniques, including k-means clustering, to partition the vector dataset into clusters or cells, and construct inverted file structures to optimize the organization of vectors within these clusters. During query execution, a fundamental trade-off between accuracy and speed is managed through the ‘nprobe’ parameter which instructs the algorithm on the number of clusters to search. This design enables the algorithm to begin with clusters closest to the query in the vector space and systematically extend its search to clusters located farther from the query vector.

Figure \ref{fig:Recallvsnprobe} shows how the accuracy (recall@1k)  vs speed (nprobe) can be tuned in MS-MARCO v1 dataset using ColBERTer’s CLS embeddings for ANN search using FAISS. As shown in Tables \ref{tab:motivationtable} and \ref{tab:motivationv2}, these models exhibit latencies ranging from tens to hundreds of milliseconds, with a significant portion of this time dedicated to ANN searches for candidate generation, aimed at achieving high recall@1K scores. For SSD based retrievals, it might be beneficial to overlap storage latencies with ANN search, but conventional retrieval pipelines typically determine the document embeddings to be accessed using the results from the ANN search.

We propose a flexible software prefetcher which utilizes the inherent characteristics of the approximate nearest neighbor search. Once a handful of the nearest clusters have been explored, it becomes increasingly likely that a significant portion of the true nearest neighbors have already been identified. This is precisely how ANN algorithms trade accuracy for speed by adjusting the ‘nprobe’ value. After visiting \(\delta\) clusters, we can generate an approximate list of document ids that might be accessed after the algorithm concludes. The assumption here is that the candidate generation process visits \(\eta=(\delta + \lambda\)) clusters where \(\delta\ll\lambda\). The prefetcher thread initiates after \(\delta\) closest clusters have been visited. The prefetcher thread simply selects the top K document ids from this approximate list and prefetches or reads the data using GDS while the algorithm visits \(\lambda\) more clusters to improve recall. Moreover, we have the option to raise the value of \(\delta\) to boost the hit rate. The prefetching budget in this context is the time it takes to visit \(\lambda\) extra clusters and can be approximated using equation (\ref{eq:eq2}). The prefetch step is measured as a percentage of the total nprobe selected and can be calculated using equation (\ref{eq:eq3}). For example in figure \ref{fig:Recallvsnprobe}, with a PrefetchStep of 10\% and \(\eta=2000\), the PrefetchBudget would approximately be 28 ms. Increasing the value of \(\eta\) or the size of the dataset, would typically increase the PrefetchBudget.

\begin{equation}
\begin{aligned}[t]
\text{PrefetchBudget} & \cong \text{ANNSearchTime}(nprobe = \eta) \\
&\quad - \text{ANNSearchTime}(nprobe = \delta) \label{eq:eq2}
\end{aligned}
\end{equation}

\begin{equation}
\text{PrefetchStep} = \frac{\delta}{\eta} \times 100\% \label{eq:eq3}
\end{equation}

The main idea here is that by prefetching the approximate list of document embeddings and overlapping this retrieval with the majority of ANN search, we only need to access a small portion of the missed documents during re-ranking. Ideally, we want the prefetcher to complete before the ANN search concludes as this allows us to completely hide the prefetcher latency. For a single query system, this is not a problem since the SSD bandwidth and the prefetch budget is often sufficient. Increasing the number of simultaneous/batch queries can demand higher bandwidth and at a certain batch threshold, it is possible for the prefetching time from storage to surpass the prefetching budget, resulting in observable prefetching latency along the critical path. We can determine the maximum query batch size to prevent this leakage by using equation \ref{eq:eq4}.

\begin{equation}
\text{Query Batch threshold} = \frac{BW_{SSD} \cdot \text{PrefetchBudget}}{\text{Data size per query}} \label{eq:eq4}
\end{equation}

\subsection{Early re-ranking and score aggregation}

We introduce an early re-ranking stage in our prefetcher which applies the MaxSim similarity kernel immediately after the embeddings are retrieved. When operating under the query batch threshold, this step allows the prefetcher to further reduce the query latency in the critical path. We can combine the results from the prefetcher thread and the main query thread by sorting the documents according to their similarity score and generating the final ranked list of documents. Of course, all this extra work is only beneficial if we achieve high prefetcher hit rate. We discuss bandwidth efficient solutions to increase the query batch threshold in the next section. 

\subsection{Bandwidth efficient solutions using partial re-ranking}

\begin{figure}
    \centering
    \includegraphics[width=0.9\linewidth]{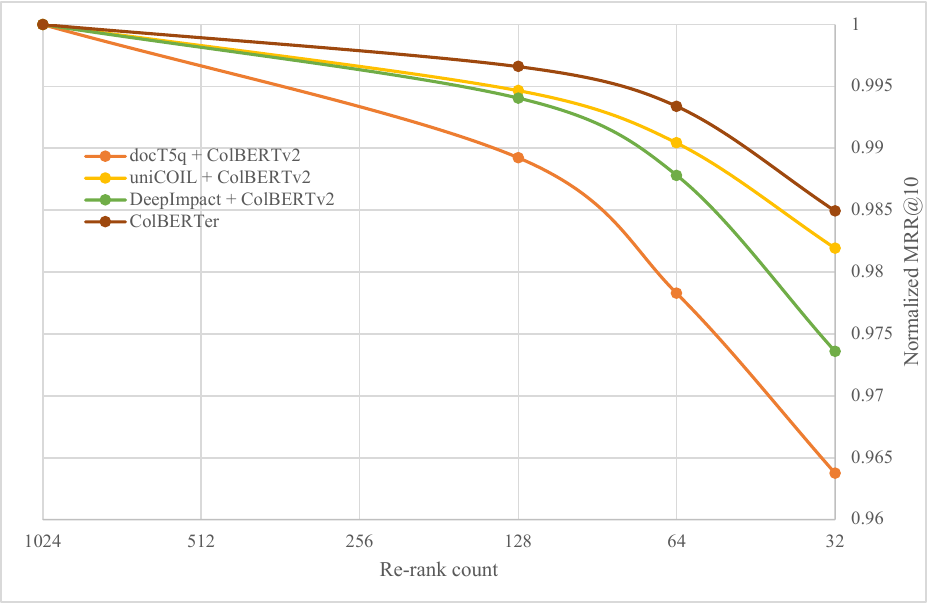}
    \caption{Normalized MRR@10 vs re-rank count with ColBERTv2, ColBERTer in MS-MARCO v1}
    \label{fig:MRRvsRerank}
\end{figure}

Our motivation behind these studies arises from the recent improvement in impact based and single-vector retrievers \cite{dai2019contextaware, lin2021brief}. Neural IR models re-rank a list of 1000 or more candidates to generate ranked list of documents. However, with the improvements in impact and single-vector retrievers, there is a potential that we can achieve approximately the same retrieval score without re-ranking the entire list of candidates. Given that these candidate generators already place relevant documents higher up in the candidate list, we can choose to employ the MaxSim operator for re-ranking a small subset of highly ranked candidates. We can simply aggregate the partially re-ranked list with the rest of the retrieved documents to generate the top-K ranked list. We evaluated ColBERTv2 with different first stage retrievers using the models from Pyserini \cite{Lin_etal_SIGIR2021_Pyserini}. In Figure \ref{fig:MRRvsRerank}, we normalized the MRR@10 scores that was achieved with re-ranking the top 1000 documents with each candidate generator and measured how the retrieval scores changed with different re-rank count for both ColBERTv2 and ColBERTer. For example, the graph shows that by only re-ranking the top 64-128 documents, we can maintain 99.3-99.7\% of the MRR@10 score in ColBERTer using its native candidate generator and 99.0-99.4\% of the MRR@10 score in ColBERTv2 using uniCOIL-tok \cite{lin2021brief} as the candidate generator. This translates to a 8-16x reduction in the amount of embedding data transferred per query which effectively allows us to increase the query batch threshold in equation (\ref{eq:eq4}). 

\section{Evaluation}

Our evaluation attempts to answer these research questions:
\begin{enumerate}[itemsep=0.5pt] 
    \item What is the effectiveness of the prefetcher? (\ref{sec:PrefetcherHitrates}) 
    \item How does ESPN affect the end-to-end query latency with different memory configurations? (\ref{sec:ExactSolutions})
    \item How does ESPN scale to large query batches? (\ref{sec:Largebatches})
    \item How does bandwidth efficient partial re-ranking improve the maximum query batch size? (\ref{sec:BWefficient})
\end{enumerate}

\subsection{Setup}
\textit{Implementation details. }We implement ESPN as an extension to ColBERTer’s codebase. We implemented the entire embedding retrieval system in C++ using Nvidia GPUDirect Storage batch APIs. We use FAISS for our ANN search and Pytorch for neural inference. Overall, ESPN constitutes roughly 600 lines of code in C++ and about 800 lines of python code. We also created a custom CUDA kernel specifically designed to parse and restructure the embedding data into the appropriate data structures essential for similarity search kernels. In our experiments, this restructuring kernel provided much better performance than multiple calls to cudaMemcpyDeviceToDevice. We modeled our prefetcher on top of FAISS and implemented the prefetcher thread with the GIL unlocked. 

\textit{Datasets. }We evaluated our system on the MS-MARCO-v1 dataset \cite{bajaj2018ms} and MS-MARCO-v2 dataset \cite{msmarcov2} on the development set queries. In the MS-MARCO v1 dataset, we use faiss with ivfflat index which had a size of 4.3 GB and in MS-MARCO v2 dataset, we use faiss with ivfpq (m=128, nbits =8) which had a size of 17.5 GB. In ESPN, these index were cached in system memory where as the re-ranking embedding tables of sizes 16.8 GB and 255.4 GB were stored in SSDs.

\textit{Systems and model}.  In both the dataset, we utilize the publically available ColBERTer model with different memory configurations. We test the retrieval system using mmap, swap space and ESPN with different memory configurations. In the v1 and v2 datasets, the total number of cells (ncell) in the ANN index was chosen to be \(2^{15}\)  and \(2^{16}\) respectively. The value of ANN nprobe was selected to be 3000 and 160 for MS-MARCO v1 and v2 respectively. The prefetch step was chosen to be at 10\% and 30\% respectively. We selected these nprobe values to strike a balance between accuracy and speed, aiming to achieve similar total latencies and retrieval scores as reported by the authors in Table \ref{tab:motivationtable}.

\textit{Hardware}. We implemented and evaluated the retrieval pipeline using an Ubuntu 20.04 machine with Intel Xeon W-2255 CPU with 10 physical cores, 256 GB of DDR4 memory and 32 GB configured for swap space. For our neural inference and MaxSim similarity kernel, we used a Nvidia A5000 GPU with 24 GB of GDDR6 memory. The storage based experiments were performed with a Samsung PM983 SSD with PCIe 3.0 interface. 

\textit{Latency Measurements. } We measure the average latency of all the queries in the dataset, including the time required for query encoding, ANN search, prefetching, early re-ranking, and final re-ranking, to measure the end-to-end retrieval latency. We measure the latency on a relatively idle machine which runs a memory limiting process in the background. To simulate different memory configurations, we employed cgroups \cite{Kerrisk_2021} to restrict the memory allocation of the python retrieval process. Simultaneously, we ran a separate process which allocated the necessary memory offset and locked its allocation using the mlock command \cite{Kerrisk_mlock}. This step was important for experiments with mmap as cgroups did not limit the size of the page cache. We made use of the madvise command for mmap with MADV\_RANDOM flag to provide hints to the OS about the expected access patterns of the query.

\subsection{Prefetcher Results} \label{sec:PrefetcherHitrates} 

\begin{figure} [h]
    \centering
    \includegraphics[width=1\linewidth]{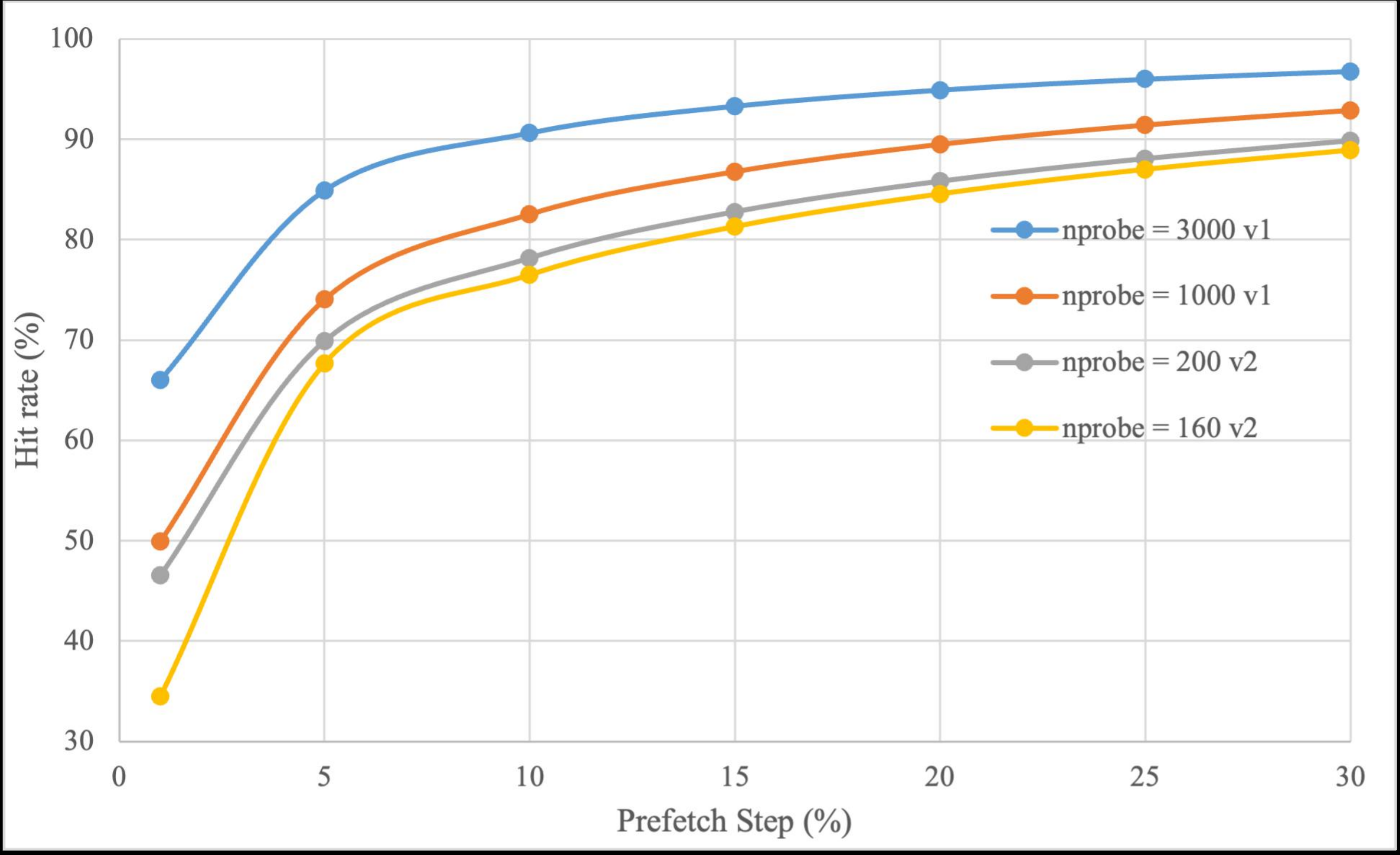}
    \setlength{\belowcaptionskip}{-10pt}
    \caption{Prefetcher hit rate vs Prefetch Step . v1 = MS-MARCO v1, v2 = MS-MARCO v2 dataset.}
    \label{fig:HitRate}
\end{figure}
\captionsetup[figure]{skip=1pt} 

In Figure \ref{fig:HitRate}, we present the hit rate achieved by the prefetcher with varying prefetching steps, expressed as a percentage of the total nprobe. Specifically, for v1, we evaluated the total nprobe values of 1000 and 3000, while for v2, we used 160 and 200. In the case of the v1 dataset, the average hit rate demonstrates substantial growth, surging from 74\% and 85\% upon visiting just 5\% of the total clusters to exceeding 90\% when exploring 30\% of the clusters. Similarly, within the v2 dataset, we observe a notable increase, with the average hit rate climbing from 68\% and 70\% at the 5\% cluster mark to a significant 89\% and 90\% after reaching 30\% of the clusters. These results show that we can prefetch the majority of the document embeddings  accurately and only access a small portion of the embeddings in the critical path of the query.

\subsection{End-to-end exact retrieval solutions } \label{sec:ExactSolutions}

\begin{table}[h]
    \centering
    \caption{End-to-end Query Latency (ms) with Different Memory Configurations in MS MARCO v1. * suggests the embedding table was cached in memory.}
    \resizebox{\columnwidth}{!}{%
    \begin{tabular}{lccccc}
        \toprule
        \textbf{Retrieval method/Memory } & 10 GB & 15 GB & 20 GB & 25 GB & 30 GB \\
        \midrule
        Mmap & 183.2  & 142.9  & 86.3  & 56.4  & 53.4  \\
        Virtual memory (+swap space) & 76.0  & 65.0  & 59.4  & 52.2  &  45.6 *  \\
        ESPN (GDS) & 54.2  & 54.0  & 52.6  & 52.5  & 52.9  \\
        ESPN (GDS + prefetcher @ 10\%) & 46.5  & 47.2  & 46.8  & 46.5  & 46.6  \\
        \bottomrule
    \end{tabular}
    }
    \label{tab:Queryv1}
\end{table}

\begin{table}[h]
    \centering
    \caption{End-to-end Query Latency (ms) with Different Memory Configurations in MS MARCO v2.}
    \resizebox{\columnwidth}{!}{%
    \begin{tabular}{lcccc}
        \toprule
        \textbf{Retrieval method/Memory } & 32 GB & 64 GB & 128 GB & 256 GB \\
        \midrule
        Mmap & 271.5 & 236.1 & 198.7 & 188.6 \\
        Virtual memory (+swap space) & OOM & OOM & OOM & OOM \\
        ESPN (GDS) & 91.2 & 90.7 & 90.02 & 90.05 \\
        ESPN (GDS + prefetcher @ 30\%) & 85.8 & 85.2 & 84.9 & 85.7 \\
        \bottomrule
    \end{tabular}
    }
    \label{tab:Queryv2}
\end{table}
Table \ref{tab:Queryv1} and \ref{tab:Queryv2} show the average end-to-end query latency with different memory configuration for MS-MARCO v1 and v2 respectively using the publicly available ColBERTer model. We achieved comparable retrieval scores as reported in Table \ref{tab:motivationtable} and \ref{tab:motivationv2} in all of the experiments in this section. Mmap based experiments only memory maps the re-ranking embedding tables to the virtual memory. When the size of the dataset is small and the index can fit in the memory, mmap appears to have low overheads because it can cache everything in the system memory after the first encounter. However, when the index size exceeds the memory capacity, the software overhead of mmap becomes apparent. This is visible when we limit the process to 10 GB of physical memory in Table \ref{tab:Queryv1}. The total index size of MS-MARCO v2 exceeded our machine’s physical memory capacity, resulting in substantial overheads when applying mmap in all the memory setups specified in Table 5. ESPN is \(3.1-3.9\times\) faster than mmap in the end-to-end query latency. Utilizing the swap space also comes with its own set of overheads and limitations. The OS brings in 8 pages per page fault when utilizing the swap space which lowers the overhead when compared to mmap.  However, this method is not scalable to larger datasets that do not fit in the memory + swap space size (Table \ref{tab:Queryv2}) and is detrimental to the SSD as it needs to constantly swap a lot of data in the swap file.

Incorporating GDS allows us to reduce the software overheads but still introduces SSD based latency in the critical path. Table \ref{tab:Queryv1} shows that ESPN with GDS and the prefetcher can achieve near memory (cached) level of query latency without storing any of the BOW embeddings in memory. In Table \ref{tab:Queryv2}, ESPN provides the most practical approximation to a cached solution as the average critical embedding access latency was \(\leq1ms\) in both the datasets. Overall, the end-to-end latency using ESPN is competitive (\(1.02\times\)) with the fully memory based solution while only storing 6 - 19\% of the total retrieval indices in memory. The average single thread query throughput using ESPN is approximately equal to the memory based solution.
This allows us to reduce the memory requirements by \(5-16\times\) depending on the quantization applied to the ANN index. These optimizations allow multi-vector models to have an index memory requirement that is only \(1.6\times\) the index memory requirement of BM25. These techniques can be used with any off-the-shelf multi-vector models to reduce their memory requirements.

\subsection{Scaling ESPN to large query batch sizes} \label{sec:Largebatches}
The difference between memory based approach and SSD based solution can be narrowed down to the embedding access latency to the GPU memory in the critical path of the pipeline. 
If the access latency is close to memory based approach, we can maintain similar query throughput for the given batch size. We simulate large query batch sizes by assuming that sufficient hardware resources (CPU/GPU cores) are available with the storage and memory bandwidth fixed, and approximating the prefetch budget to remain constant. The prefetch budget typically increases at higher batch sizes as the contention for hardware resources increases. To model this setup, we retrieve an increasing amount of embedding data with a constant prefetch budget to simulate the results with large batch queries.

\begin{figure}[h]
    \centering
    \includegraphics[width=0.9\linewidth]{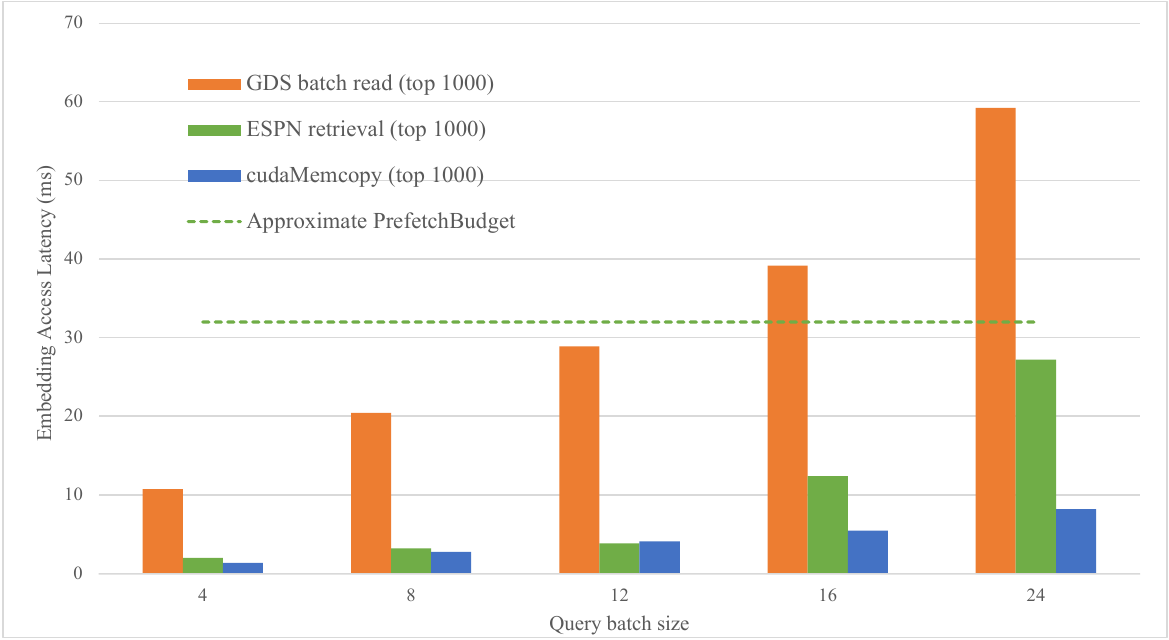}
    \caption{Exact solution: Query Batch size vs Critical path embedding access latency in MS-MARCO v1 dataset with prefetch step = 10\% }
    \label{fig:ESPNbatch}
\end{figure}

\begin{figure}[h]
    \centering
    \includegraphics[width=0.9\linewidth]{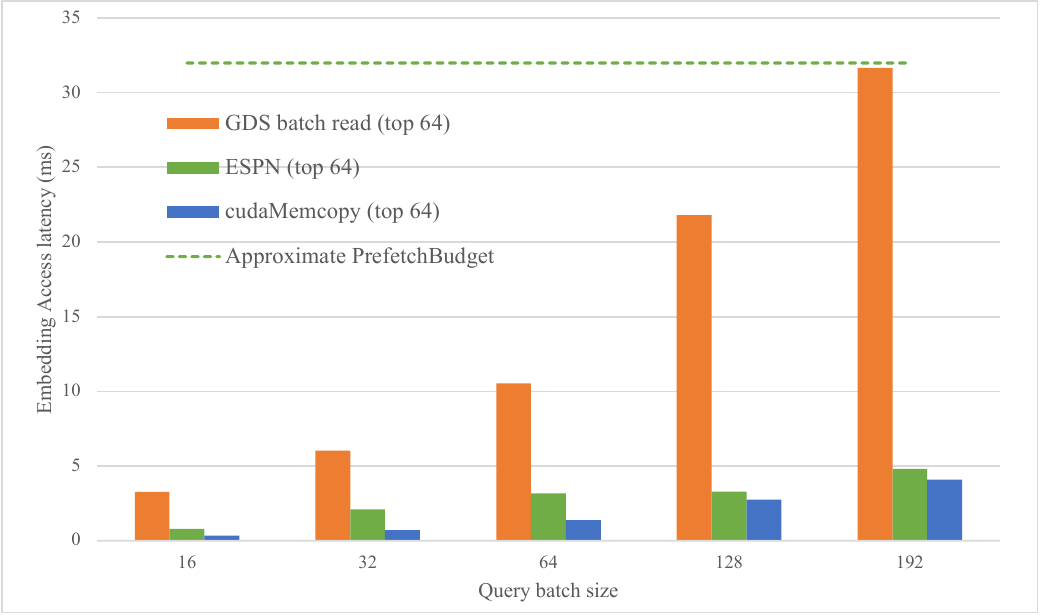}
    \caption{Bandwidth efficient solution: Query Batch size vs Critical path embedding access latency in MS-MARCO v1 dataset with prefetch step = 10\% }
    \label{fig:BWEfficiency}
\end{figure} 

\begin{figure}[h]
    \centering
    \includegraphics[width=0.9\linewidth]{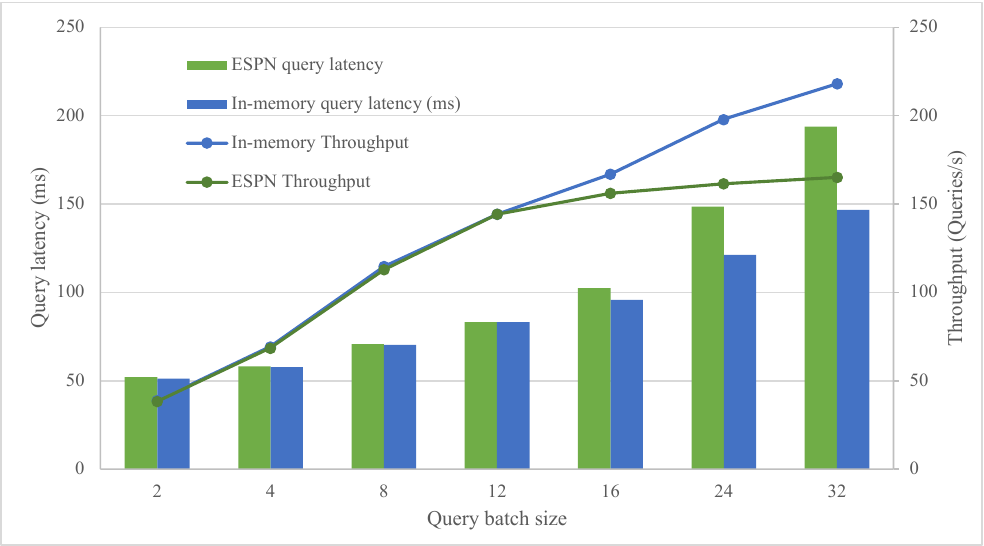}
    \caption{Exact solution: Batch Query latency and throughput in MS-MARCO v1 dataset with prefetch step = 10\% }
    \label{fig:batchlatency}
\end{figure}

In Figure \ref{fig:ESPNbatch} and \ref{fig:BWEfficiency}, we compute the approximate PrefetchBudget for the v1 dataset using prefetch step of 10\%. To make fair comparisons with GPUDirect Storage, we measure the SSD latency using Nvidia’s gdsio tool \cite{nvidiagds}. As the batch size grows, there is a linear increase in the number of embeddings that need to be retrieved from the storage. In these experiments, we model each query to access 1000 BOW document embeddings from storage. Figure \ref{fig:ESPNbatch} shows how SSD based solution (GDS) introduces \(7.2\times\) higher embedding access latency compared to DRAM. ESPN can maintain near memory levels of access latency as long as the prefetcher operates under the prefetching budget. Operating within this budget, ESPN accesses a small fraction of the total embeddings in the critical path, enabling it to minimize the number of retrievals and maintain low access latencies. We model the end-to-end query latency and the throughput of ESPN with these results and compare it with the memory based solution in Figure \ref{fig:batchlatency}. 

Using a PCIe gen 3.0 SSD, Figure \ref{fig:ESPNbatch} and \ref{fig:batchlatency} shows that ESPN is competitive with DRAM based embedding retrieval up to a batch size of 12. Within this limit, ESPN improves GDS based embedding retrieval by \(3.9-6.4\times\) depending on the batch size. Newer SSDs with PCIe gen 4.0 should increase the total random bandwidth by \(2\times\) and increase this limit to around 24 according to equation (\ref{eq:eq4}). Once the batch size surpasses this threshold, the prefetcher’s retrieval latency extends into the critical path. In our current design, the main thread waits for the prefetcher thread before it starts retrieving the remaining data from storage. We discuss how we can further improve upon this in the future work section.

\subsection{Bandwidth efficient partial re-ranking solution} \label{sec:BWefficient}

The partial re-ranking solution significantly reduces the number of the embeddings being retrieved per query, and, therefore, the prefetcher consistently operates below the prefetch budget for larger batch sizes. We have essentially decreased the data size per query in the equation (\ref{eq:eq4}) which allows us to increase the query batch size limit per SSD. In figure \ref{fig:BWEfficiency}, each query accesses only the top 64 document embeddings instead of accessing the top 1000. By reducing the bandwidth requirement by almost \(16\times\), ESPN, using the SSD configuration, is competitive with memory based embedding retrieval up to a batch size of 192. The bandwidth efficient partial re-ranking solution trades a marginal reduction (0.7\%) in the MRR@10 retrieval score to increase the scalability of SSD based retrievals to larger batch sizes.

\section{Related work}

Considerable research efforts have focused on optimizing computational and memory efficiency of training and inference of neural systems \cite{Aminabadi2022DeepSpeedIE, rajbhandari2022deepspeed-moe, 10.1145/3600006.3613165, pope2022efficiently}. Dedicated hardware accelerators , such as tensor cores in GPUs and TPUs, have significantly enhanced the computational efficiency of neural inference \cite{9623445, 10.1145/3579371.3589350}. Furthermore, distillation, pruning, and quantization techniques have managed to reduce the number of model parameters and improve inference speed \cite{hinton2015distilling, han2016deep}. Additionally, algorithmic redesign and the integration of fused kernels from FlashAttention have yielded massive improvement in transformer inference speed while reducing memory consumption \cite{dao2022flashattention}. SSD based systems have been studied for training and neural recommendation inference to improve memory requirements \cite{bae2021flashneuron, 10.1145/3445814.3446763}. DiskANN proposes a novel graph based indexing algorithm which enables moving ANN index to SSDs and minimizes latencies by reducing access to disk resident index \cite{subramanya2019diskann}. Similarly, SPANN proposes an IVF based solution which offloads majority of ANN index to storage \cite{ChenW21}.
While DiskANN and SPANN only focuses on ANN search, we focus on the entire Neural IR pipeline and offload the substantial re-ranking index to SSDs. We also build an efficient pipeline employing a novel flexible prefetcher and early re-ranking to reduce the impact of SSD latencies. We take inspiration from many of these recent advancements and apply them towards Neural IR.

There has been substantial work to train neural models to learn and improve embeddings and representations for data which can be used for search \cite{xiong2021approximate, 10.1145/3488560.3498443, liu2021pretrained, gao-callan-2022-unsupervised, qu-etal-2021-rocketqa, DBLP:journals/corr/abs-2104-08051}. In the realm of neural IR, the cost of inference is reduced by using offline indexing but this process inflates the index size and memory requirements for efficient inference. To reduce the index memory footprint, techniques such as Product Quantization and PCA is typically used to compress the vectors which can be lossy and reduce retrieval quality \cite{5432202, MACKIEWICZ1993303}.
New precision formats for deep learning have been introduced which can further reduce the memory footprint \cite{10.1145/3579371.3589351, micikevicius2022fp8}.
We can apply these techniques in conjunction with the strategies introduced in this paper; quantizing the ANN index size reduces the memory requirements and compressing the re-ranking embedding index reduces the data size per query in equation (\ref{eq:eq4}) which improves access latency. ANN search with Faiss can also be accelerated using GPUs \cite{DBLP:journals/corr/JohnsonDJ17}. However, competitive speedups require the ANN index to be cached in GPU HBM which becomes increasingly difficult and expensive for large datasets. It is possible to aggressively compress the ANN index to fit in the limited GPU memory at the cost of lower retrieval scores. Employing ESPN with GPU based ANN search could require rewriting GPU kernels which is a potential future research direction.

\section{Limitations and Future work}

We built our embedding retrieval solution on top of the relatively new Nvidia GPUDirect storage, which can have some limitations in its current version. For example, using an IO size of 4096 bytes we were not able to fully saturate the SSD bandwidth using the batch APIs. We also ran into some minor issues like having to reregister the GPU buffers after each IO to get high throughput from SSD which can be expensive. We expect a lot of these issues to be solved as the GDS system matures with newer releases. 

One of the intricacies in multi vector systems involves variations in the size of each BOW embedding. The size of the BOW embedding varies based on the document's word/token count, ranging from as small as 2KB to as large as 10KB in the datasets tested. When an embedding's size slightly exceeds the I/O limit, it necessitates the retrieval of an extra block, despite only a fraction being relevant to the current query. A potential solution to address these issues is to employ a larger I/O size (e.g., 8KB or more) and pack multiple embeddings within the boundaries, enabling the retrieval of multiple BOW embeddings per I/O operation. This could further increase the throughput from SSDs and minimize any inefficiencies. While we focus our current work on ANN based candidate generation, we also plan to use these techniques with lexical retrievers in future work. A logical next step to improve our design is to take inspiration from systems like DiskANN, SPANN and offload the majority of the candidate generation index to SSDs as well. In our experiments, we use a single SSD for embedding retrieval which can have a limited bandwidth. GDS supports RAID 0 configurations where we could potentially combine multiple SSDs to fully saturate the PCIe bandwidth and increase the throughput from storage. 
Using larger number of storage devices with the newer PCIe and CXL interconnects will continue providing increasing device bandwidths, and as a result we expect these solutions to sustain high query throughputs \cite{9912551, sharma2023introduction}.
We leave these potential optimizations as future work.

\section {Conclusion}
 The considerable index sizes of Multi-vector IR models often present the most substantial challenge when implementing these systems at scale. In this paper, we propose Embedding from Storage Pipelined Network (ESPN) which offloads the entire re-ranking embeddings to SSDs and reduces the memory requirements up to \(16\times\). We introduce a novel ANN based prefetching mechanism with hit rates exceeding 90\%, improving SSD based retrieval up to \(6.4\times\), and maintaining near memory levels of query latency and throughput. We also explore bandwidth efficient solutions which allows ESPN to scale to larger query batches. In summary, ESPN reduces the index memory footprint and accelerates storage access allowing efficient and scalable inference of Neural Multi-Vector Information Retrieval.



\bibliography{espn}
\bibliographystyle{mlsys2024}

%


\end{document}